\documentclass[onecolumn,amsmath,showpacs,amssymb,showkeys]{revtex4}
\usepackage{amsfonts}
\usepackage{graphicx}
\usepackage{epsfig}
\usepackage{bm}

\begin{document}

\title{\large\bf The polarized Bjorken sum rule analysis: revised
\footnote{
Presented by V.~L. Khandramai at the XXVIth International
Conference \& School ``Foundations \& Advances in Nonlinear
Science'',  Minsk, Belarus, September 2012.}
}

\author{V.~L. Khandramai$^{\dag}$, O.~P. Solovtsova$^{\dag}$, O.~V. Teryaev$^\ddag$}

\affiliation{\small {\it $^\dag$ Gomel State Technical University, 246746 Gomel, Belarus} \\
{\small {\it $^\ddag$ Joint Institute for Nuclear Research, 141980
Dubna, Russia}}}

\begin{abstract}
We present progress in the QCD analysis of the Bjorken sum rule
at low momentum transfers. We study asymptotic structure of the
perturbative QCD expansion at low $Q^2$ scales based on analysis
of recent accurate data on the Bjorken sum rule and available
now four-loop expression for the coefficient-function
$C_{Bj}(Q^2)$. We demonstrate that the standard perturbative
series for $C_{Bj}(Q^2)$ gives a hint to its asymptotic nature
manifesting itself in the region $Q^2 \lesssim 1$~GeV$^2$. It is
confirmed by the considered integral model for the perturbative
QCD correction. We extract a value of higher-twist $\mu_4$
coefficient and study the interplay between higher orders and
higher-twist contributions.
Results of other approaches to the description
of Bjorken sum rule data are discussed.
\end{abstract}



\maketitle

\section{Introduction}
Perturbation theory (PT), supplemented by the renormalization
group method and the renormalization procedure, is a main tool
of QCD calculations. Conventional perturbative expansions for
hadronic observables are the power series in the strong coupling
$\alpha_{\rm s}$. Owing to the property of asymptotic freedom in
QCD, the perturbative description becomes more reliable at high
energy region. On the contrary, at low momentum transfer, $Q^2
\lesssim 1-2$~GeV$^2$, the use of the perturbative power series
may be questionable. Therefore, it is important to know at least
a few terms of PT series to estimate theoretical uncertainties
associated with a truncation of the perturbative expansion.
Currently, the perturbative QCD analysis is aimed at the
perturbative calculations of new terms of PT series.
Only recently up to the fourth-order perturbative approximation
for some physical processes became available
\cite{Baikov:2008jh,Baikov:2010je}.
The number of these processes includes the fundamental sum rule
for polarized deep inelastic scattering which is the Bjorken sum
rule $\Gamma_1^{p-n}(Q^2)$ \cite{Bj} defined as a difference
between the first moments of the proton $g_1^{\rm p}$ and the
neutron $g_1^{\rm n}$ spin structure functions.
The Bjorken sum rule has been measured in polarized deep inelastic
lepton scattering at SLAC, CERN, DESY \cite{2,3,4,5,6,7}. At low
$Q^2$, in a $Q^2$-range from $0.05<Q^2<3.2$~GeV$^2$,
high-precision data on the $\Gamma_1^{p-n}$ has been presented by
the Jefferson Lab~\cite{Deur:2004ti} and at $3.0$~GeV$^2$ by the
COMPASS collaboration~\cite{Alekseev:2010hc}.

At low momentum scales, the QCD analysis of the Bjorken sum rule
includes  both the perturbative and non-perturbative
higher-twist (HT) components related to each other. It is clear
that the reliability of extracting information on the HT effects
is connected to the indeterminacy in the description of the PT
series. As new perturbative correction of an order of
$\alpha_{\rm s}^4$ to the Bjorken sum rule became
available~\cite{Baikov:2010je}, the opportunity for new
researches opened.

In this report, we present results of the four-loop QCD analysis
of the Bjorken sum rule at low momentum scales, continuing our
investigations started in
Refs.~\cite{Pasechnik:2009yc,Khandramai:2011zd}. We consider the
features of the four-loop PT series and  the interplay between
the higher PT order and the HT contributions. We discuss results
of application of other approaches which were applied to the
description of low $Q^2$ Jefferson Lab data.

\section{Standard approach}

Away from the large $Q^2$ limit, the polarized Bjorken sum rule
is given by a double series in powers of $\alpha_{\rm s}$
and in powers of $1/Q^2$
(nonperturbative HT corrections):
\begin{eqnarray} \label{Eq:2-PT-Bj-HT}
\Gamma^{p-n}_1(Q^2)=\frac{|g_A|}{6}\biggl[1-\Delta_{\rm Bj}^{\rm
PT}(Q^2)\biggr]+\sum_{i=2}^{\infty}\frac{\mu_{2i}}{Q^{2i-2}}\,,
\end{eqnarray}
where $|g_A|=1.2701\pm0.0025$~\cite{Beringer:1900zz} is the
nucleon axial charge, $\mu_{4},\, \mu_{6}, ~\dots~ $ are the HT
coefficients; $\Delta_{\rm Bj}^{\rm PT}(Q^2)$ is the
perturbative correction which is defined by the
coefficient-function $C_{Bj}(Q^2)$, $\Delta_{\rm Bj}^{\rm
PT}(Q^2)\equiv 1-C_{\rm Bj}(Q^2)$.
Note that until very recently $\Delta_{\rm Bj}^{\rm PT}(Q^2)$
has been known up to the third order in $\alpha_{\rm s}$. The
corresponding expression was used in many studies (see, e.g.,
Refs.~\cite{Ellis:1994py,Pasechnik:2009yc,MSS-Bj,Penida}), in
particular, to extract a value of $\alpha_{\rm s}$ from
experimental data. Comparison of these values with other
accurate $\alpha_{\rm s}$ values, such as those obtained from
the $\tau$-lepton and the $Z$-boson into hadrons width decays,
is an important test of the consistency of QCD
\cite{Beringer:1900zz,Bethke:2012jm}.

\subsection{Four-loop analysis}

\begin{figure}[bth]
\begin{center}
         \begin{minipage}[bht]{0.5\textwidth}
\centerline{\includegraphics[width=7.8cm]{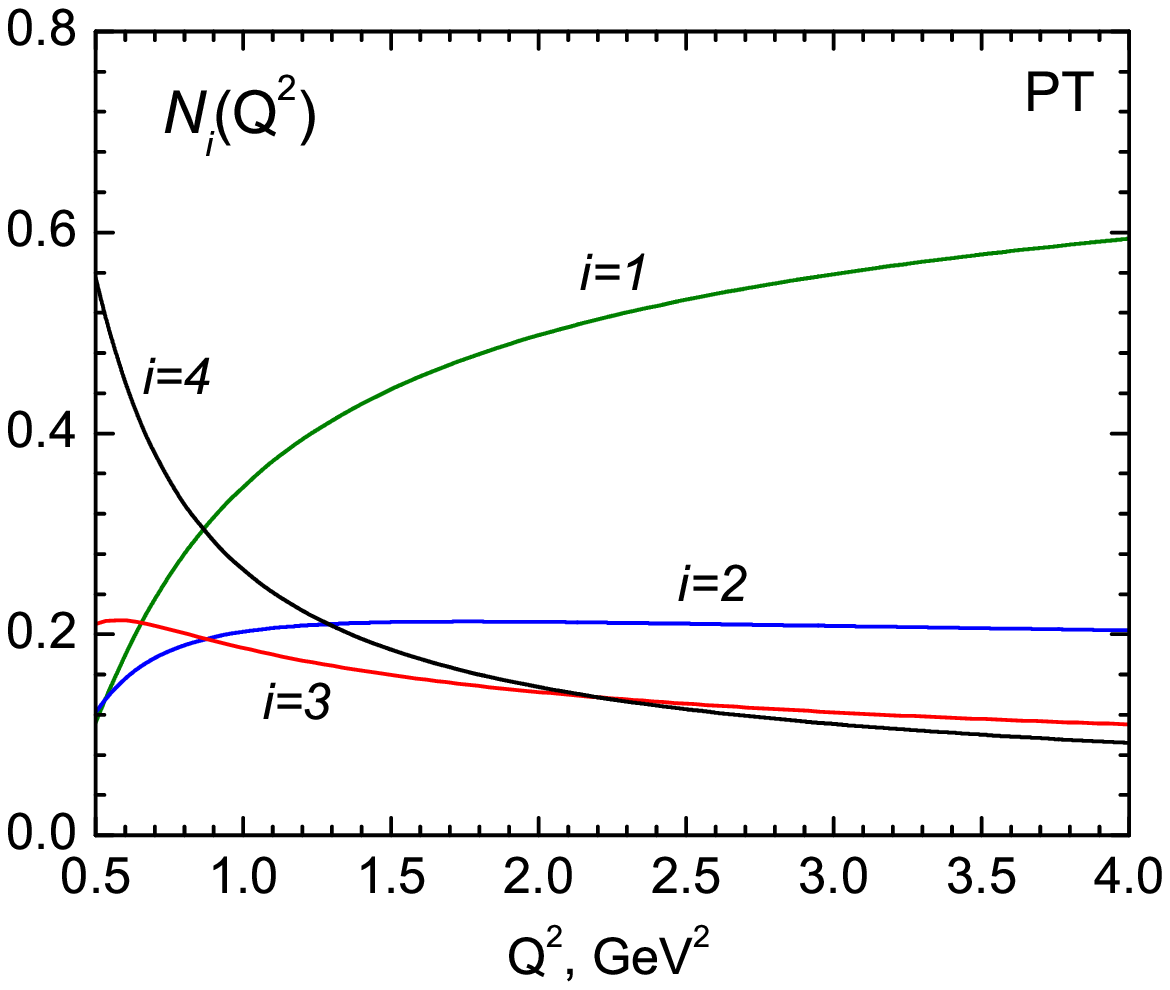}}
         \end{minipage}
         \phantom{}\hspace{-0.2cm}%
     \begin{minipage}[bht]{0.5\textwidth}
\centerline{\includegraphics[width=7.8cm]{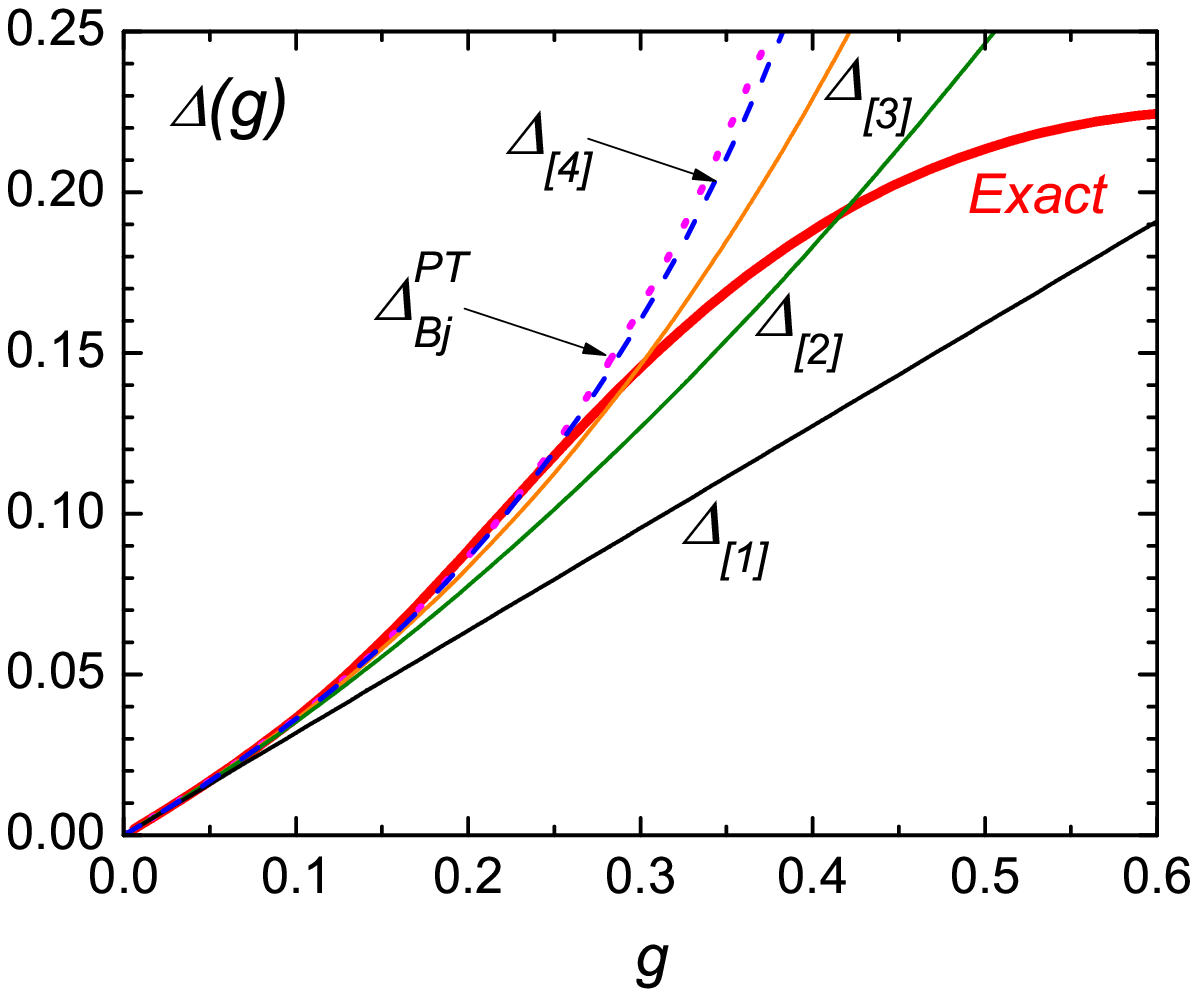}}
    \end{minipage}
  \begin{minipage}[t]{0.45\textwidth}
\caption{The relative contribution ${\rm N}_i(Q^2)=c_i
\alpha_{\rm s}^i(Q^2)/\Delta_{\rm Bj}(Q^2)$
for the $i$-th term of series (\ref{Eq:3-Delta_PT-Bj})
as a function of the $Q^2$.}
\label{Fig:1}
        \end{minipage}%
\phantom{}\hspace{0.9cm}%
     \begin{minipage}[t]{0.47\textwidth}
\caption{Comparison of the 1-, 2-, 3-, and 4-term approximants of
series (\ref{Eq:5-Model_Num}) with the exact result
(\ref{Eq:4-IntModel}) and with N$^3$LO Bjorken sum rule series
(\ref{Eq:3-Delta_PT-Bj}).} \label{Fig:2}
    \end{minipage}
     \end{center} \end{figure}

At the four-loop (N$^3$LO) level, the perturbative QCD
correction to the Bjorken sum rule in the case of massless
quarks in the $\overline{\rm MS}$ renormalization scheme for
$f=3$ light quarks flavors reads as \cite{Baikov:2010je}
\begin{eqnarray}
\label{Eq:3-Delta_PT-Bj} \Delta_{\rm Bj}^{\rm
PT}=0.318\,\alpha_{\rm s}+0.363\,\alpha_{\rm
s}^2+0.652\,\alpha_{\rm s}^3+\,1.804\,\alpha_{\rm s}^4\,.
\end{eqnarray}
We use the PT running coupling ${\alpha}_{\rm s}(Q^2)$ obtained
by integration of the renormalization group equation with the
four-loop $\beta$-function  (see Ref.~\cite{Khandramai:2011zd}
for additional details).

Let us discuss the convergence properties of the PT power series
(\ref{Eq:3-Delta_PT-Bj}) at low momentum transfers.
Figure~\ref{Fig:1} shows the relative contribution of the $i$-th
term of this series as a function of $Q^2$. As can be seen from
this figure, in the region of small $Q^2 \lesssim 1$ GeV$^2$ the
dominant contribution comes from the $\alpha_{\rm s}^4$-term.
This may be considered as an indication of the transition of PT
series to the asymptotic regime and one can estimate the value
$\alpha_{\rm s}(1\,{\rm GeV}^2)\simeq0.5$ as a critical. In the
region $Q^2>2$~GeV$^2$ the situation changes -- the major
contribution comes from the first and second terms.

To specify discussed above convergence properties of series
(\ref{Eq:3-Delta_PT-Bj}), we consider the exact example
following from the integral model~\cite{Kazakov:1980rd}  (see
Ref.~\cite{Shirkov:2012zm} for more details):
\begin{equation} \label{Eq:4-IntModel}
\displaystyle C(g)=\frac{1}{\sqrt{\pi}}\int^{\infty}_{-\infty}
\,e^{-x^2(1-\frac{\sqrt{g}}4\,x)^2}\,dx = 1+\frac{1}{\sqrt{\pi}}\sum_{k=1}^{\infty}
C_k g^k \,, \quad C_k= \frac{\Gamma(2k+1/2)}{4^k\,\Gamma(k+1)}\,.
\end{equation}
Using this expression, one can build series which is  close to the
series (\ref{Eq:3-Delta_PT-Bj}):
\begin{equation} \label{Eq:5-Model_Num}
\Delta(g)=\frac{16}{3\pi}\left[C(g)-1\right]\, = \,
0.318\,g+0.348\,g^2+0.718\,g^3+2.188\,g^4+\dots~.
\end{equation}

In Fig.~\ref{Fig:2} we compare the finite sum approximations
$\Delta_{[n]}(g)={\sum}^{n} c_i g^i$, $n=1,\dots,4\,$  both with
the exact result and with the Bjorken series
(\ref{Eq:3-Delta_PT-Bj}). This figure demonstrates the closeness
of the 4-term approximation (dashed curve) and the N$^3$LO
approximation (\ref{Eq:3-Delta_PT-Bj}) (dotted curve). It is
necessary to note, that 1-, 2-, and 3-order approximations to
$\Delta(g)$ and $\Delta_{\rm Bj}$ practically coincide with each
other. This figure shows that the 2-term approximant is good up
only to $g=0.15-0.20$ and the 3-term one up to $g\sim0.33$ while
the 4-term sum starts to deviate from exact $C(g)$-curve at
$g\sim 0.27\,.$ We interpret this observation as asymptotic
structure manifestation.
This model confirms that the asymptotic structure of the pQCD
expansions manifests. Therefore, the application of standard PT
greater than N$^2$LO approximation can not allows to extract
accurate information in the low-energy domain.

\subsection{Higher twist contribution}

We expand our consideration, including the HT part which is
presented in expression (\ref{Eq:2-PT-Bj-HT}). In
Table~{\ref{T1} we show our results for values of the
coefficient $\mu_4$ (the errors are statistical only) fitted to
the low $Q^2$ data~\cite{Deur:2004ti,Alekseev:2010hc} in
different PT orders. One can see that $\mu_4$-values extracted
changes rather strongly between different PT orders. The
absolute value of $\mu_4$ decreases with the order of PT and
just at N$^3$LO becomes compatible to zero. It can be
interpreted as a manifestation of duality between higher orders
and HT (see Ref.~\cite{Narison:2009ag}).
%
\begin{table}[tbh]
\centering
\caption{Results of  $\mu_4$-extraction from the data on the Bjorken sum rule in
different PT orders.}
\hphantom{}
\label{T1}
\begin{tabular}{ccccc}
\hline
PT order  & ~~LO & ~~NLO  & ~N$^2$LO & ~N$^3$LO
   \\ \hline
$\mu_4$,~GeV$^2$ & $~-0.037\pm0.003$ & $~-0.025\pm0.004$ & $~-0.012\pm0.006$  & $~~0.005\pm0.008$
   \\ \hline
\end{tabular}
\end{table}
Note that a value extracted in the leading-order (LO) is
consistent with a value $\mu_4=-0.047\pm0.025$~GeV$^2$ presented
in Ref.~\cite{Ross:1993gb} as well as with  a value
$\mu_4=-0.028\pm0.019$~GeV$^2$ obtained from the
next-to-leading-order (NLO) fit based on the $x$-dependent
structure functions data \cite{Sidorov:2006vu} .

\section{Other approaches}

As shown above, use of the conventional PT series even with the
HT component does not allow to describe the Jefferson Lab data
down to the infrared region (see Fig.~\ref{Fig:3}). Let us
consider some other approaches which are more successful in this
direction.

\subsection{APT}

In Ref.~\cite{Shirkov:1997wi} the conventional method of the
renormalization group improvement of the perturbative expansions
was modified by requiring K$\ddot {\rm a}$llen-Lehmann
analyticity, which reflects the principle of causality. In the
framework of this approach~\cite{Shirkov:1997wi} called as the
analytic perturbation theory (APT) the ghost pole and
corresponding branch points, which appear in higher PT orders,
are absent (see, e.g., Ref.~\cite{MSS-time}). As the moments of
the structure functions should be analytic functions in the
complex $Q^2$ plane with a cut along the negative real axis (see
Ref.~\cite{JLD-00}  for more details), the standard description
of the $\Gamma_1^{p-n}(Q^2)$ violates analytic properties due to
the unphysical singularities of perturbative  running coupling.
On the other hand, the APT support these analytic properties
(see Ref.~\cite{Shirkov:2007} as review). The perturbative
correction to the Bjorken sum rule can be written in the form of
a spectral representation~\cite{MSS-Bj} and at four-loop level
is
\begin{equation}\label{Eq:6-Delta_APT-Bj}
\Delta_{\rm Bj}^{\rm
APT}=0.318\,{\cal{A}}_{1}+0.363\,{\cal{A}}_{2}+0.652\,{\cal{A}}_{3}+\,1.804\,{\cal{A}}_{4}\,.
\end{equation}
The power coefficients in this expression are the same as in
series (\ref{Eq:3-Delta_PT-Bj}), and the functions
${\cal{A}}_{k}(Q^2)$ are defined through the spectral density
$\varrho_k(\sigma)\equiv{\rm Im}\left[\alpha_{\rm s}^{k}
(-\sigma-i\epsilon)\right]$ by the spectral integral (see
Ref.~\cite{Khandramai:2011zd} for additional details).
At large momentum transfers, all the functions
${\cal{A}}_{k}(Q^2)$ become proportional to the $k$-th power of
the usual perturbative coupling $\left[\alpha_{\rm
s}(Q^2)\right]^k$ and the expansion (\ref{Eq:6-Delta_APT-Bj})
reduces to the power series (\ref{Eq:3-Delta_PT-Bj}). However,
at small enough
$Q \lesssim  1 - 2$~GeV the properties of the
non-power expansion (\ref{Eq:6-Delta_APT-Bj}) become
considerably different from the PT power series
(\ref{Eq:3-Delta_PT-Bj}).

Figure~\ref{Fig:3} shows the results of $\mu_4$-fits in
different PT orders both in the PT and APT approaches. From this
figure follows that in the framework APT including
$\mu_4$-coefficient allows one to move further down to $Q^2 \sim
0.1$~GeV$^2$~\cite{Khandramai:2011zd} in description of the
Jefferson Lab data. It is important to note that the APT leads
to higher-loops stability of the HT-extraction:
$\mu_4=-0.044\pm0.004$~GeV$^2$ in all loop approximations.

\begin{figure}[bth]
\begin{center}
         \begin{minipage}[bht]{0.5\textwidth}
\centerline{\includegraphics[width=8.2cm]{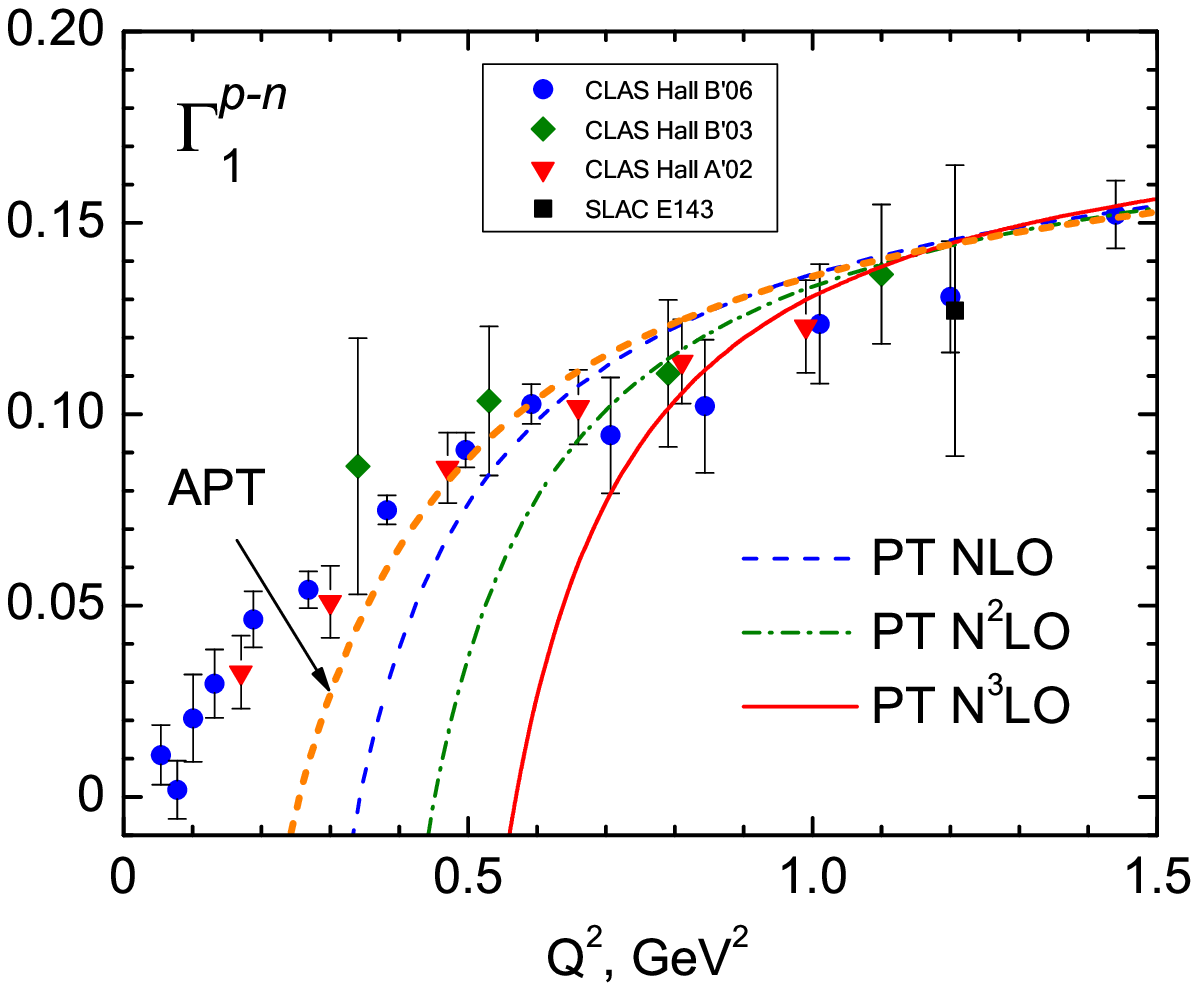}}
         \end{minipage}
         \phantom{}\hspace{-0.2cm}%
     \begin{minipage}[bht]{0.5\textwidth}
\centerline{\includegraphics[width=8.2cm]{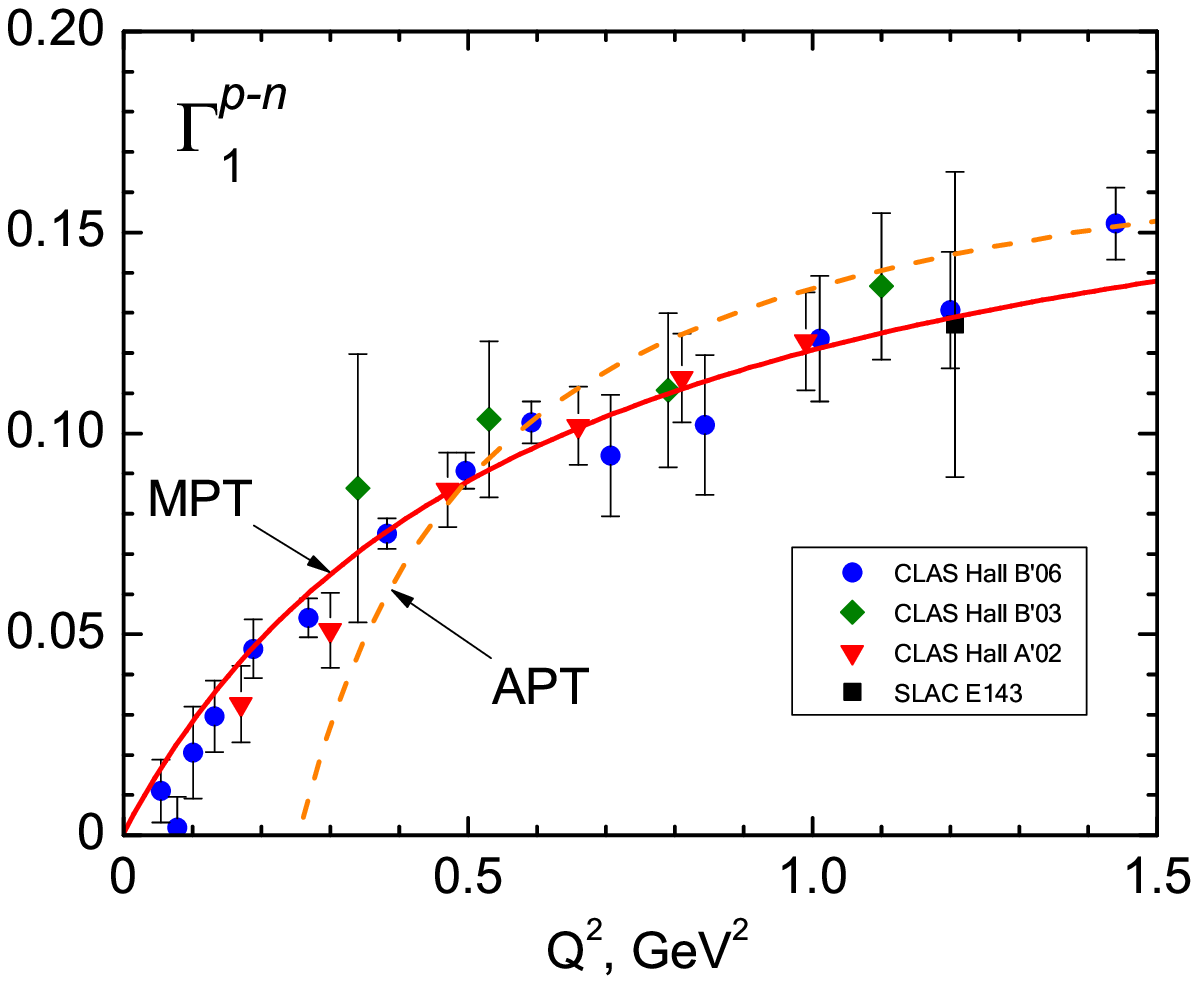}}
    \end{minipage}
  \begin{minipage}[t]{0.45\textwidth}
\caption{The $\mu_4$-fit of the Jefferson Lab data in different
orders of the PT and the all-order APT expansions.}
\label{Fig:3}
        \end{minipage}%
\phantom{}\hspace{0.9cm}%
     \begin{minipage}[t]{0.47\textwidth}
\caption{The MPT fitting of the Jefferson Lab data.} \label{Fig:4}
    \end{minipage}
     \end{center} \end{figure}

\subsection{MPT}

In context of data on the $\Gamma_1^{p-n}$ analysis it should be
noted the MPT approach \cite{Shirkov:2012ux}. Basis of the MPT
is a simple idea to change the usual logarithm in the expression
for the QCD running coupling, $1/\ln(Q^2/\Lambda^2)\,$, that is
singular in the infrared region, on the  ``long logarithm''
$\ln(\xi+Q^2/\Lambda^2)\,$, where parameter $\xi$ corresponds to
the ``effective gluonic mass'' $m_{gl}=\sqrt{\xi}\,\Lambda$:
\begin{equation}\label{Eq:7-a1-mpt}
{\cal A}_{1, MPT}(Q^2)=\frac1{\beta_0\,
\ln(\xi+Q^2/\Lambda^2)}.
\end{equation}

Result of  MPT application can be interpreted as a transition to
the new momentum-transfer scale both in the PT and the HT
contributions: $ \Delta_{\rm Bj}^{\rm MPT}\sim\Delta_{\rm
Bj}^{\rm PT}(Q^2+m_{gl}^2)$ and
$~\mu_4/Q^2\sim{\mu_4}/{\left(Q^2+m^2_{ht}\right)} $. The MPT
curve is shown in Fig.~\ref{Fig:4} as solid line (preliminary
result Ref.~\cite{KhSh12}). From this figure one can see that
the MPT approach allows us to describe the Bjorken sum rule data
down the infrared region  $Q^2 \to0 $.

Note that very close description of the Jefferson Lab data was
obtained in Ref.~\cite{Kotikov:2012eq}, where was used a
modified expression for the coefficient-function $C_{\rm Bj}$ in
a combination with
 ``frozen'' behavior of running coupling ${\alpha}_{\rm s}(Q^2)$.

\section{Summary}

We have presented the QCD analysis of the Bjorken sum rule up to
four-loop level in $\alpha_{\rm s}$ at low momentum transfers.
It has been shown that the asymptotic nature of the perturbative
series manifested itself at the four-loop level in the region
$Q^2 \lesssim 1$ GeV$^2$. Therefore, at low $Q^2$ the inclusion
of the four-loop term does not improve the precision of
theoretical predictions. It was confirmed by the considered
integral model for the perturbative QCD correction.

Using the recent low $Q^2$-data from the Jefferson Lab and
COMPASS experiments, we have extracted a value of higher-twist
$\mu_4$ coefficient and have shown the interplay between higher
orders and higher-twist contributions. Results of other
approaches to the description of Bjorken sum rule data have
discussed.

\acknowledgments
It is a pleasure to thank D.~V.~Shirkov for useful advice and
stimulating discussions as well as S.~V.~Mikhailov and
A.~V.~Kotikov for interest to this work. This work was supported
in part by the BelRFBR-JINR grant,~F12D-002 and RFBR
grant,~11-01-00182.


\end{document}